\documentclass[twocolumn,tightenlines,aps,preprintnumbers,superscriptaddress,showpacs,floatfix]{revtex4}
\usepackage{graphicx}
\usepackage{epsf}
\usepackage{amssymb}
\usepackage{amsmath,amssymb,textcomp,array}

\newcommand{\tr}{\text{tr}\,}
\newcommand{\sd}{\text{sd}\,}

\begin{document}

\preprint{}

\title{Chaos and Noise}
\author{Temple He}
\affiliation{Department of Physics, Stanford University, 382 Via
  Pueblo Mall, Stanford CA 94305}
\affiliation{Department of Physics, Harvard University, 17 Oxford Street, 
Cambridge MA 02138}  
\author{Salman Habib}
\affiliation{Theoretical Division, MS B285, Los Alamos National
  Laboratory, Los Alamos, NM 87545}
\affiliation{High Energy Physics Division \& Mathematics and Computer
  Science Division, Argonne National Laboratory, 9700 South Cass
  Avenue, Lemont, IL 60439}

\date{\today}

\begin{abstract}

  Simple dynamical systems -- with a small number of degrees of
  freedom -- can behave in a complex manner due to the presence of
  chaos. Such systems are most often (idealized) limiting cases of
  more realistic situations. Isolating a small number of dynamical
  degrees of freedom in a realistically coupled system generically
  yields reduced equations with terms that can have a stochastic
  interpretation.  In situations where both noise and chaos can
  potentially exist, it is not immediately obvious how Lyapunov
  exponents, key to characterizing chaos, should be properly
  defined. In this paper, we show how to do this in a class of
  well-defined noise-driven dynamical systems, derived from an
  underlying Hamiltonian model.

\end{abstract}


\maketitle

\section{Introduction}

The characterization of deterministic chaos, particularly in
Hamiltonian systems, is now well-established, and essentially consists
of computing and understanding the Lyapunov spectrum of the dynamical
system~\cite{eckmann,gaspard}. In the vast majority of experiments,
however, Hamiltonian systems are only an idealization, and it is not
always clear -- given a phenomenological description -- what role
dynamical chaos might actually play in the physics, let alone be
certain about how to characterize it.

A somewhat similar situation also exists in the modeling of complex
systems, where often one wishes to separate `slow' and `fast' degrees
of freedom in such a way that the fast degrees of freedom can be
viewed as a stochastic forcing term, often called `noise'~\cite{ngvk,
  rwz, kj}. As emphasized by Zwanzig~\cite{zwanzig1}, a key problem is
that depending on the nature of the approximations one chooses to
make, the resulting noise terms in the stochastic dynamical equations
(Langevin equations), can end up being very different.

It is therefore easy to appreciate that when both noise and chaos are
combined, the situation is doubly complicated. Nevertheless, such
situations are not only common, but also of significant interest in
many applications. These include such diverse areas as noise-induced
chaos~\cite{crutchfield, gao99a, hwang00, nic},
ecology~\cite{ellner05}, galactic dynamics~\cite{galdyn}, and the
quantum-classical transition~\cite{bhj,hjs}.

The purpose of this paper is to consider the interaction of noise and
chaos in a dynamical model where the basic physical and
approximation-related issues can be separately understood. This will
not only help clarify the nature of some previous disagreements in the
literature but also provide a more well-founded notion of the
(maximal) Lyapunov exponent (LE) in noisy systems, at least within a
specific, well-defined, context.

The overall picture adopted here assumes the existence of a complex
Hamltonian system, where one focuses on a small set of relevant
degrees of freedom, deriving effective equations for their evolution.
Because the full system is specified, issues of how to define chaos
and noise, and their interaction, are easier to clarify. Discussions
of the possible difficulties and choices of definitions in the context
of a more phenomenological setting can be found in
Refs.~\cite{vulp95,gao99}.

In this paper, we consider Hamiltonian models for Brownian motion as
the relevant archetype. In these models, a dynamical system is coupled
to a heat bath, modeled by an ensemble of non-interacting harmonic
oscillators. An analogous model was first described by
Rubin~\cite{rjr}, and the basic notion was later elucidated by a
number of authors, including Ford, Kac, and Mazur~\cite{fkm},
Zwanzig~\cite{zwanzig2}, and, as better known in the quantum context,
by Caldeira and Leggett~\cite{cl}.  Although these models are by no
means completely general, they provide an excellent basis for
addressing conceptual problems.

In adopting this view we nevertheless wish to be clear in what
circumstances analyses such as ours are meant to apply: Our results
will not be directly relevant to heuristic models of complex systems
that are not based on a first-principles analysis, nor are they
intended to apply in the strong noise limit, where the effects of
chaos can be washed out by the effects of the noise drive.

As a consequence of the self-consistent nature of our analysis, the
(nonequilibrium) fluctuation-dissipation theorem will ensure that
system trajectories explore a unique distribution at late times, that
defined by thermal equilibrium, $f_{eq}\sim\exp\{-\beta H\}$, where
$H$ is the system Hamiltonian and $\beta=1/k_BT$ is proportional to
the inverse temperature. Thus the LE we consider is
averaged over the canonical distribution, in contrast to a constant
energy hypersurface in the non-noisy case, when the system is not
coupled to a heat bath. The lower the temperature of the heat bath,
the smaller the noise strength, and the smaller the energy
fluctuations. In the singular limit of vanishing coupling to the heat
bath, any phase space distribution that is a function of the system
Hamiltonian is invariant, and there is no longer the notion of a
single unique distribution that can be used to study trajectories in
the late-time limit. However, in this case, as a consequence of the
ergodic theorem, constant energy hypersurfaces are fully explored by
long-time trajectories, and, as is familiar, the LEs
become functions of the energy.

At this point it is important to distinguish between `external' and
`internal' noise. The case of external noise arises when the noise
drive is unaffected by the system evolution, while internal noise
refers to the fluctuations of a complex system coupling to its own
slow modes; in this case, the noise may be affected by the system
evolution, both in terms of its intrinsic statistical properties and
in terms of the particular noise realization associated with a
background trajectory (even if the statistics are unaffected). We
discuss both of these cases below.

Finally, it is important to consider various limits, such as the
overdamped (strong-coupling) limit and the weak-noise (low
temperature) limit. Note that these limits are independent, but can
define physical timescales relevant to our analysis. In the examples
considered here, the system relaxation time is independent of
temperature or noise strength and depends only on the damping
coefficient, in turn set by the inverse of the system-bath coupling
strength. The convergence timescales for the LE are typically very
long, significantly longer than typical thermal relaxation timescales,
consequently the initial phase of the evolution is unimportant. Thus,
for our purposes, the relevant dynamics of the system is that related
to the exploration of a thermal equilibrium state as mentioned above.
In the cases we study, the equilibrium is established by interaction
with a heat bath (equilibrium for a canonical ensemble, as given by
the static solution of a Fokker-Planck equation), satisfying the
fluctuation-dissipation relation. Energy fluctuations are bounded as
long as the system potential is asymptotically confining. For the most
part we will consider the low noise case but discuss both the weak and
strong coupling limits. The strong coupling limit is singular and care
is needed with the analysis in order to avoid incorrect results.

The rest of the paper is organized as follows. We begin in
Section~\ref{sec_two} with a short review of the derivation of the
relevant Langevin equations, emphasizing issues that will be taken up
in later sections. We will then consider the case of the Langevin
equation with the noise treated entirely as an external perturbation
(Section~\ref{sec_three}), define the LE, and proceed to show --
without making the overdamped approximation -- that when the system is
one-dimensional, the addition of noise cannot lead to a positive
LE. This changes the result of Ref.~\cite{broeck} who claimed that in
this case, the LE could be positive (noise-induced chaos) or negative.
Extending to higher dimensions, we show that in this case, the LE can
indeed be positive or negative, in contrast to Ref.~\cite{loreto}, who
claimed that the LE is always negative. Having completed the analysis
for the external noise case, we then extend our analysis to systems
with internal noise (Section~\ref{int_noise}). Examples of numerical
results that agree with our conclusions can be found in
Refs.~\cite{bhj,hjs,galdyn}. Finally, we discuss some open questions
and possible future research directions related to chaos and noise in
our conclusion.

For notational consistency we will write all our vectors in boldface
and all our matrices in scripts, and the components of vectors and
matrices will be in normal font, i.e. $v_1$ is the first component of
the vector $\mathbf v$ and $M_{11}$ is the top-left entry of the
matrix $\mathcal M$.

\section{Independent Oscillator Model for Brownian Motion}
\label{sec_two}

As discussed in the Introduction, for the most part, stochastic
differential equations appear in the modeling of physical systems
primarily in a phenomenological context. Because of the uncontrolled
approximations inherent in how these equations are often written down,
it is not obviously apparent how to think about the interaction of
noise and chaos in a systematic fashion, as both of these can be very
sensitive to choices made in the modeling process, which may or may
not be entirely self-consistent, or even physically correct.

There are, nevertheless, several examples of areas where stochastic
equations can be derived in a more or less controlled fashion (based
on assumptions such as timescale separation), starting from an initial
Hamiltonian formulation. These include the (multiplicative noise)
Langevin description of the Landau equation in plasma
physics~\cite{landau1}, the description of Brownian motion based on a
Hamiltonian system coupled to a large set of independent
oscillators~\cite{rjr,fkm,zwanzig2,cl}, and, more generally, equations
derived using the Mori-Zwanzig projection operator
technique~\cite{rwz_proj,mori,grabert}.

To fix ideas, we present here a short derivation of Brownian motion
using the independent oscillator model, following
Zwanzig~\cite{zwanzig1}. The full (positive-definite) Hamiltonian, of
a system interacting  with a ``bath'' of oscillators is taken to be:
\begin{equation}\label{full_ham}
	H=\frac{1}{2}p^2+V(x)+\frac{1}{2}\sum_j\left[p_j^2+
          \omega_j^2\left(q_j-a_j(x)\right)^2\right],      
\end{equation}
where the system coordinates are $(x,p)$ and the smooth functions
$a_j(x)$ describe possibly nonlinear couplings of the system to the
oscillators. Note that there is an assumption here that no degree of
freedom of the bath is strongly perturbed by the system of
interest; this justifies treating the $q_j$ effectively in a
linearized approximation. The distribution of the frequencies,
$\omega_j$, as well as the choice of the function $a_j(x)$ controls
the noise memory as discussed below.

To simplify matters, we first consider a linear coupling to the
oscillators, i.e., we set $a_j(x)=(\gamma_j/\omega_j^2)x$. Writing
down Hamilton's equations starting from Eq.~(\ref{full_ham}), formally
solving the equations of motion for the oscillators treating
$\gamma_jx(t)$ as time-dependent external force, and substituting this
solution into the  equations for the system variables, we obtain the
formal Langevin equation, 
\begin{equation}\label{langevin}
	\dot{p}(t)=-V'(x)-\int_0^tK_N(t-s)p(s)+F_N(t),
\end{equation}
where the damping kernel is
\begin{equation}\label{kernel}
	K_N(t-s)\equiv\sum_j\frac{\gamma_j^2}{\omega_j^2}\cos\omega_j(t-s),
\end{equation}
and the `noise' force is
\begin{eqnarray}\label{noise}
	F_N(t)=&&\sum_j\gamma_j\left[q_j(0)-
          \frac{x(0)}{\omega_j^2}\gamma_j\right]\cos\omega_jt\nonumber\\  
&&+\sum_j\gamma_jp_j(0)\frac{\sin\omega_jt}{\omega_jt}.
\end{eqnarray}
The damping kernel, or memory function, is in general not Markovian
and is determined entirely by the coupling constants and oscillator
frequencies. In contrast, $F_N(t)$ is determined entirely by the
initial conditions. If these are known precisely then $F_N(t)$ is not
a noise term. However, if the initial conditions are prescribed in a
statistical manner, then the situation is different. Suppose that the
statistical ensemble of initial conditions is such that the first
moments vanish, i.e.,  
\begin{eqnarray}
	\langle p_j(0)\rangle_0&=&0, \nonumber\\
	\langle q_j(0)-\frac{\gamma_j}{\omega_j^2}x(0) \rangle_0&=&0,
\end{eqnarray}
then, it is immediate from Eq.~(\ref{noise}) that $\langle F_N(t)
\rangle_0=0.$ Now, if the second moments of the initial ensemble are,  
\begin{eqnarray}
  \langle p_j(0)p_k(0)\rangle_0&=&k_BT\delta_{jk}, \nonumber\\
  \langle [q_j(0)-\frac{\gamma_j}{\omega_j^2}x(0)] 
 [q_k(0)-\frac{\gamma_k}{\omega_k^2}x(0)]\rangle_0&=&
\frac{k_BT}{\omega_j^2}\delta_{jk},
\end{eqnarray}
then,
\begin{equation}\label{fdt}
	\langle F_N(t) F_N(t')\rangle_0=k_BTK_N(t-t'), 
\end{equation}
which is a generalized fluctuation-dissipation theorem. An initial
ensemble that satisfies these conditions is one in which the heat bath
is in thermal equilibrium with respect to the system, whereas the
system variables are allowed to have an arbitrary distribution:
\begin{equation}\label{ic_dist}
f(t=0)\sim f^{sys}_0(x,p)\exp(-\beta H_{bath}),
\end{equation}
where $H_{bath}$ represents the Hamiltonian of the oscillators and
the system-bath couplings, i.e., the term containing the summation in 
Eq.~(\ref{full_ham}). In general, the noise is not Markovian;
however, in the special limiting case of a large number of
oscillators, $N$, following a Debye distribution (i.e., the spectral
distribution $g(\omega)=3\omega^2/\omega_D^2$ for $\omega<\omega_D$, 
and $g(\omega)=0$ for $\omega>\omega_D$), and assuming that the system 
momentum varies slowly on timescales set by the inverse Debye cutoff
$\tau_D=1/\omega_D$, the kernel $K_N(t-s)$ can be approximated as a
delta function, and the Langevin equation takes on the more familiar
form
\begin{equation}\label{langevin_debye}
\dot{p}(t)=-V'(x)-\lambda p(t)+F_N(t),
\end{equation}
where the noise is `white' due to the sharpness of the memory kernel,
and it is Gaussian thanks to the quadratic nature of $H_{bath}$ in
Eq.~(\ref{ic_dist}). Here, the damping coefficient is
$\lambda=3\pi\gamma^2/2\omega_D^2$, where we set $\gamma_j\rightarrow 
\gamma/\sqrt{N}$ when taking the limit of a large number of 
oscillators. The fluctuation-dissipation relation (\ref{fdt}) now becomes 
\begin{equation}\label{fdt_cont}
\langle F_N(t) F_N(t')\rangle_0=2\lambda k_BT\delta (t-t').
\end{equation}
It is important to point out here that the simplified derivation given
above can be considerably sharpened: The system trajectories can be
rigorously shown to converge to the solutions of a stochastic problem
both in the weak~\cite{kupferman04} and strong sense~\cite{ariel09}.

Returning to the case of an arbitrary nonlinear coupling specified by
$a_j(x)$, the above analysis goes through essentially unchanged, but
with a more complicated memory kernel associated with multiplicative
noise (albeit, still white and Gaussian in the case of the Debye
spectrum) that continues to satisfy the fluctuation-dissipation
theorem. The presence of multiplicative noise can cause qualitatively
new dynamical effects because the noise amplitude depends on the
system variables. Such effects include modifications of the
equilibration rate~\cite{mult_rel1, mult_rel2} and the existence of
long-time tails in transport theory~\cite{zwanzig_tails}. The basic
procedure described in this paper can be extended to the case of
multiplicative noise as long as the system is being analyzed in the
asymptotic late-time limit, i.e., on timescales much longer than the
relaxation time.

Although this class of models is relatively simple, it can be extended
in interesting directions by changing the system potential or by
manipulating the spectral distribution of bath oscillators. Examples
of such extensions include studies of escape problems and tests of
transition state theory~\cite{pollak89, ariel07} and Hamiltonian
models leading to fractional kinetics~\cite{kupferman04b}.

In the models discussed above, the fluctuation-dissipation theorem
follows as a consequence of how we chose the initial condition for the
bath oscillators. This choice corresponds to running many copies of
the system drawn from some initial distribution but, in which, for
each realization of the initial condition, the oscillator bath is in
thermal equilbrium with respect to $x(0)$. The system following the
Langevin equation (\ref{langevin_debye}) will be driven at late times
to the thermal equilbrium distribution $f_{eq}\sim\exp{(-\beta H_{sys})}$, where
$H_{sys}$ corresponds to the first two terms of the
full Hamiltonian $H$, as specified in Eq.~(\ref{full_ham}). (This
result can be most easily derived by considering the associated
Fokker-Planck equation for the phase space
distribution~\cite{risken}.) Consequently, time-averaged quantities
exist ($\bar{f}_{\tau}(x,p)=1/\tau\int_0^{\tau}f(x,p;t)dt$), and are
stable in the limit $\tau\rightarrow\infty$. We note that we are
explicitly not considering systems with an external time-drive in
which case there is, in principle, no equilibrium state.

Note also that the Langevin equation derived here is inherently
second-order in time (or, more generally, an even number of
first-order equations). In the limit of strong coupling (large
$\lambda$), the velocity can be eliminated as a `fast' variable and,
in leading approximation, a first-order stochastic equation may be
derived~\cite{gardiner}. This limit is singular, however, and care
must be taken in applying it in different situations.

Finally, we turn to a discussion of how to define LEs
for the system considered here. In principle, there is no problem,
since given the full Hamiltonian and prescribed initial conditions, we
can consider a small perturbation of the initial conditions around
some fiducial trajectory of the (full) coupled system. The summary of
this procedure is as follows. For a 2$n$-dimensional dynamical system
governed by a set of evolution equations, $d{\mathbf z}/dt={\mathbf
 F}({\mathbf z},t)$, where ${\mathbf z}= (z_1, z_2, \cdots, z_{2n})^T$
(similarly for ${\mathbf F}$), consider (i) a fiducial trajectory,
${\mathbf z_0}(t)$, (ii) define deviations from it via ${\mathbf
 Z}={\mathbf z}-{\mathbf z}_0$, and (iii) linearize the original set
of equations, yielding,
\begin{equation}
\frac{d{\mathbf Z}}{dt}={\mathcal{DF}}({\mathbf z}_0,t)\cdot{\mathbf Z},
\label{lin_eqn}
\end{equation}
where $\mathcal {DF}$ is the $2n\times 2n$ Jacobian matrix. The
tangent map $\mathcal Q({\mathbf z}_0(t),t)$ is found by integrating
the linearized equations along the fiducial trajectory; $\mathcal
Q({\mathbf z}_0(t),t)$ evolves the initial variables ${\mathbf
 Z}_{in}$ via ${\mathbf Z}(t)=\mathcal Q(t) {\mathbf Z}_{in}$. Define
the $2n\times 2n$ matrix $\mathcal L$ as $\mathcal
L=\lim_{t\rightarrow\infty}(\mathcal Q\tilde{\mathcal Q})^{1/2t}$,
where ${\tilde{\mathcal Q}}$ is the matrix transpose of $\mathcal
Q$. The LEs are then given by the logarithm of the eigenvalues of
$\mathcal L$.

If we are going to view the oscillator variables in a statistical
sense, however, then we are interested only in the Lyapunov spectrum, and in
particular the maximal LE, of the system variable $(x,p)$ given an
ensemble of initial conditions for the oscillator variables (we will
throughout this paper refer to the maximal LE simply as the LE). Note
that because the system evolution is actively coupled to the
environment variables, for each realization of a set of oscillator
initial conditions, the system trajectory will itself be
different. Moreover, as given explicitly in the definition of the
noise term Eq.~(\ref{noise}), any perturbations in the initial
condition, $x(0)$, as required for defining the associated LE, would
automatically change the realization of the noise force. This is an
expected consequence of any systematic procedure as applied to a
coupled Hamiltonian system. In our independent oscillator model, the
noise is therefore a particularly simple case of {\em internal} noise.

\section{Systems with External Noise}
\label{sec_three}

We first consider the case of systems subjected to external noise. In
this case, one assumes that the noise realizations are completely
independent of the initial conditions of the system variables. This
would be the straightforward interpretation of
Eqs.~(\ref{langevin_debye}) and (\ref{fdt_cont}) if we began our
analysis with these two equations and the actual nature of their
derivation was not specified. We will consider internal noise in the
next section.

We consider an $n$-dimensional system in coordinate space, with the
system trajectory writen as ${\mathbf x}(t) \equiv
(x_1(t),\ldots,x_n(t))^T$. The (Gaussian, additive white
noise) Langevin equation we consider here is essentially
Eq.~(\ref{langevin_debye}) written in a slightly generalized form,
\begin{equation}\label{langeq}
  \ddot{\mathbf x}(t) + \gamma\dot{\mathbf x}(t) = -{\mathbf\nabla}
U({\mathbf x}(t)) + \sqrt{\mathcal D}\,dW_{ext}(t), 
\end{equation} 
where $dW_{ext}(t)$ is the external noise term, with normalization set
by $(dW_{ext})^2=dt$, $U(\mathbf x(t))$ the
potential, and $\mathcal D$ an $n \times n$ diagonal matrix, with
the $k$th diagonal entry, denoted $D_k$, being the noise intensity
along $x_k$. Furthermore, $\gamma$ is the damping coefficient
($\gamma>0$), a time-independent scalar. Let ${\mathbf x}_0(t)$ be the
fiducial trajectory, and write ${\mathbf x}(t) = {\mathbf x}_0(t)
+ {\mathbf\Delta}(t)$, where ${\mathbf\Delta}(t) \equiv
(\Delta_1(t),\ldots,\Delta_n(t))^T$. Upon linearization,
\begin{equation}\label{1dlin}
	\ddot{{\mathbf\Delta}}(t) + \gamma\dot{\mathbf\Delta}(t) =
        -{\mathcal A}(t)\cdot{\mathbf\Delta}(t), 
\end{equation}
where $\mathcal A(t) \equiv (\partial_{x_i,x_j}U({\mathbf x}_0(t)))$, the
Jacobian of ${\mathbf\nabla} U(\mathbf x_0(t))$. Notice that the
noise term has disappeared in the linearized equation as we are
considering it to be external. This means that any result we obtain
will formally resemble that of the noise-free system; the only
difference is that the noise may alter the fiducial trajectory
$\mathbf x_0(t)$ and hence the time average of quantities that depend
on it. 

Now let
\begin{equation}\label{zdef}
	{\mathbf z}(t) = {\mathbf\Delta}(t)e^{\gamma t/2},
\end{equation}
and substitute into the linearized equation \eqref{1dlin}, to yield
\begin{equation}\label{zlin}
	\ddot{\mathbf z}(t) - {\mathcal B}(t)\cdot{\mathbf z}(t) = 0,
        \quad\text{where} \quad {\mathcal B}(t) = \frac{\gamma^2}{4} -
        {\mathcal A}(t). 
\end{equation}
Using a more canonical notation,
\begin{equation}\label{vdef}
	{\mathbf v}_1(t) = {\mathbf z}(t), \quad {\mathbf v}_2(t) =
        \dot{\mathbf z}(t), 
\end{equation}
Eq.~\ref{zlin} can be written as
\begin{align}\label{veq}
	&\begin{pmatrix} 
  	\dot{\mathbf v}_1(t) \\ 
   	\dot{\mathbf v}_2(t)
	\end{pmatrix} = {\mathcal M}(t)\begin{pmatrix} 
  	{\mathbf v}_1(t) \\ 
   	{\mathbf v}_2(t)
	\end{pmatrix}, \nonumber\\
	\text{where}\quad &\mathcal M(t)
        \equiv \begin{pmatrix}  
  	\mathbf 0 & \mathcal I \\ 
   	\mathcal B(t) & \mathbf 0
	\end{pmatrix}.
\end{align}
Notice here $\mathcal M$ is a $2n \times 2n$ matrix. Since both
${\mathbf v}_1(t)$ and ${\mathbf v}_2(t)$ are $n$-dimensional column
vectors, our above matrix equation is in fact a system of $2n$
equations. Its formal solution is given by
\begin{align}\label{formsol}
	\begin{pmatrix}\nonumber
  	{\mathbf v}_1(t) \\ 
   	{\mathbf v}_2(t)
	\end{pmatrix} &= \bigg(\mathcal I + \int_0^t \mathcal
        M(t_1)\,dt_1 \\
&\quad + \int_0^t\int_0^{t_1}\mathcal M(t_1)\mathcal M(t_2)\,dt_2\,dt_1
\nonumber\\ 
&\quad + \cdots\bigg)  \begin{pmatrix}\nonumber
  	{\mathbf v}_1(0) \\ 
   	{\mathbf v}_2(0)
	\end{pmatrix} \\
	&\equiv  \mathcal Q(t){\mathbf v}(0),
\end{align}
where $\mathcal Q(t) = Te^{{\mathcal M}(t)}$, the time-ordered exponential
of ${\mathcal M}(t)$, and ${\mathbf v}(t) = ({\mathbf v}_1(t),{\mathbf
  v}_2(t))^T$. 

Because long-time averages exist as stated in Section~\ref{sec_two}, we
can write ${\mathcal M}(t) = \bar{\mathcal M} + {\mathcal F}_M(t)$ where
each entry of ${\mathcal F}_M(t)$ oscillates around zero. Denoting
$\bar{\mathcal Q}(t) = Te^{\bar{\mathcal M}}$, our aim is now to show that,
as far as computing the maximal LE is concerned,
we can ignore the contribution ${\mathcal F}_M(t)$ and be able to
substitute $\bar{\mathcal Q}(t)$ for ${\mathcal Q}(t)$ in
Eq.~\eqref{formsol}. To justify such a substitution, we first use
Eqs.~\eqref{zdef} and \eqref{vdef} to write Eq.~\eqref{formsol} as
\begin{eqnarray}\label{formsol2}
	\begin{pmatrix}
		{\mathbf\Delta}(t) \\
		\dot{\mathbf\Delta}(t) + \frac{\gamma}{2}{\mathbf\Delta}(t)
	\end{pmatrix} &=& e^{-\gamma t/2}{\mathcal Q}(t){\mathbf
          v}(0)\nonumber\\ 
&=& e^{-\gamma t/2}\bar{\mathcal Q}(t){\mathcal R}(t){\mathbf v}(0)
\end{eqnarray}
where we define $\mathcal R(t)$ via $\bar{\mathcal Q}(t)^{-1}\mathcal Q(t)
\equiv \mathcal R(t)$ (we assume $\bar{\mathcal Q}(t)$ is invertible). Every entry in ${\mathcal R}(t)$ grows slower
than a linear exponential; to see this, note that 
\begin{eqnarray}\label{expandM}
{\mathcal R}(t)&=&\bar{\mathcal Q}(t)^{-1} {\mathcal Q}(t)\nonumber\\
&=&Te^{{\mathcal F_M}(t)+HOT(\bar{\mathcal M},{\mathcal F_M}(t))}
\end{eqnarray}
where the terms designated as $HOT$ are given by the
Campbell-Baker-Hausdorff series. Since $\bar{\mathcal M}$ is a bounded 
constant and ${\mathcal F_M}(t)$ is a bounded oscillating function, if
we consider the RHS of the above equation as a matrix, then every
entry in the matrix grows slower than a linear exponential.

Without loss of generality, let $|\Delta_{i_0}(t)|$ be the component
of ${\mathbf\Delta}(t)$ with the largest exponential dependence, and let
$\bar Q_{i_0j_0}R_{j_0k_0}$ be the entry in $\bar{\mathcal
 Q}(t)\mathcal R(t)$ with the largest exponential
dependence. Then the maximal LE of the system is given by 
\begin{align}\label{lyapest}
	\lim_{t \to \infty}\frac{1}{t}\log \Delta_{i_0}(t) &=
        \lim_{t \to \infty}\frac{1}{t}\log\left(\sum_{j,k}\bar
          Q_{i_0j}R_{jk}v_{k}(0)e^{-\gamma t/2}\right) \nonumber\\ 
	&= \lim_{t\to\infty}\frac{1}{t}\log(\bar
        Q_{i_0j_0}R_{j_0k_0}(t)v_{k_0}(0)) - \frac{\gamma}{2} \nonumber\\ 
	&= \lim_{t\to\infty}\frac{1}{t}\log(\bar Q_{i_0j_0}v_{k_0}(0))
        \nonumber\\
        &\quad+ \lim_{t\to\infty}\frac{1}{t}\log R_{j_0k_0}(t) -
        \frac{\gamma}{2} \nonumber\\ 
	&= \lim_{t\to\infty}\frac{1}{t}\log(\bar Q_{i_0j_0}v_{k_0}(0))
        - \frac{\gamma}{2} \nonumber\\ 
	&= \lim_{t\to\infty}\frac{1}{t}\log(\bar
        Q_{i_0j_0}v_{j_0}(0)) - \frac{\gamma}{2} 
\end{align}
where in the second-to-last line we used the fact $R_{ij}(t)$ grows
slower than a linear exponential, and hence 
$\lim_{t\to\infty}(1/t)\log R_{i_0j_0}(t) =
\lim_{t\to\infty}1/t^{\epsilon(t)} = 0$, where $\epsilon(t) > 0$. In
the last line we used the fact $v_{k_0}(0)$ is just a constant term
and hence we can simply replace it with the constant term $v_{j_0}(0)$
without affecting the exponential behavior. Furthermore, since
$\bar{\mathcal Q}(t) = Te^{\mathcal{\bar{M}}} = e^{\mathcal{\bar M}t}$,
$\bar{\mathcal Q}(t)$ is a linear exponential of $t$. Thus, as
$\bar Q_{i_0j_0}R_{j_0k_0}$ is the term in $\bar{\mathcal Q}(t)\mathcal R(t)$
with the largest exponential dependence, and all the entries of
$\mathcal R(t)$ grow slower than a linear exponential, $\bar
Q_{i_0j_0}$ must be the term in $\bar{\mathcal Q}(t)$ with the largest
exponential dependence. Therefore, the final line in the expression
above is just the LE calculated had we replaced $\mathcal Q(t)$ with
$\bar{\mathcal Q}(t)$.

Using the above arguments, we can substitute $\mathcal Q(t)$ as
$\bar{\mathcal Q}(t)$ without affecting the LE. Hence we
may write 
Eq.~\eqref{formsol} as
\begin{equation}\label{avgVevol}
	\begin{pmatrix}
  	{\mathbf  v}_1(t) \\ 
   	{\mathbf v}_2(t)
	\end{pmatrix} \sim Te^{\bar{\mathcal M}}\begin{pmatrix}
		{\mathbf  v}_1(0) \\
		{\mathbf  v}_2(0)
	\end{pmatrix}.
\end{equation}
Furthermore, since $Te^{\bar{\mathcal M}} = e^{\bar{\mathcal M}t}$ as
$\bar{\mathcal M}$ is constant, we have 
\begin{equation}\label{vsol}
	\begin{pmatrix}
        {\mathbf  v}_1(t) \\ 
   	{\mathbf  v}_2(t)
	\end{pmatrix} \sim \exp\left[\begin{pmatrix} 
  	\mathbf 0 & \mathcal I \\ 
   	{\bar{\mathcal B}} & \mathbf 0 
	\end{pmatrix}t\right]\begin{pmatrix}
		{\mathbf  v}_1(0) \\
		{\mathbf  v}_2(0)
	\end{pmatrix}.
\end{equation}
Direct expansion of the matrix exponential yields
\begin{equation}\label{matexp1}
	\exp\left[	\begin{pmatrix} 
  	\mathbf 0 & \mathcal I \\ 
   	{\bar{\mathcal B}} & \mathbf 0 
	\end{pmatrix}t\right] = \begin{pmatrix} 
  	\cosh(\sqrt{\bar{\mathcal B}}t) & \sqrt{\bar{\mathcal
          B}}^{-1}\sinh(\sqrt{\bar{\mathcal B}}t) \\  
   	\sqrt{\bar{\mathcal B}}\sinh(\sqrt{\bar{\mathcal B}}t) &
        \cosh(\sqrt{\bar{\mathcal B}}t)  
	\end{pmatrix},
\end{equation}
where $\sqrt{\bar{\mathcal B}}$ is the matrix square root of $\bar{\mathcal
B}$. Substituting Eqs.~\eqref{zdef} and \eqref{vdef} into
Eq.~\eqref{vsol} and using Eq.~\eqref{matexp1}, we get
\begin{eqnarray}\label{Delta1sol}
{\mathbf \Delta}(t) &\sim& \cosh(\sqrt{\bar{\mathcal B}}t)e^{-\gamma
 t/2}{\mathbf v}_1(0) \nonumber\\
&&+ \sqrt{\bar{\mathcal B}}^{-1} \sinh(\sqrt{\bar{\mathcal
          B}}t)e^{-\gamma t/2}{\mathbf v}_2(0). 
\end{eqnarray}
To calculate $\sqrt{\bar{\mathcal B}}$, note that $\mathcal A$, given
in Eq.~\eqref{1dlin}, is a real symmetrical matrix, so $\mathcal B$
(and hence $\bar{\mathcal B}$), defined in Eq.~\eqref{zlin}, is also a
real symmetrical matrix and is hence diagonalizable with real
eigenvalues. Then we can write $\bar{\mathcal B} = \mathcal   
V\mathcal D\mathcal V^{-1}$, where $\mathcal D$ is a diagonal
matrix. It follows that $\sqrt{\bar{\mathcal B}} = \mathcal
V\sqrt{\mathcal D}\mathcal V^{-1}$, and $\sqrt{\mathcal D}$ is just
the square root of the entries along the diagonal of $\mathcal
D$. Thus, we see from Eq.~\eqref{Delta1sol} that if we want
${\mathbf\Delta}(t)$ to have exponential divergence, we require
$e^{\pm\sqrt{\bar{\mathcal B}}t}$ to have an exponential with power
greater than $\gamma/2$ as one of its terms. In particular,
\begin{equation}\label{expsqrtBt}
	e^{\pm\sqrt{\bar{\mathcal B}}t} = e^{\pm \mathcal V\sqrt{\mathcal
            D}t\mathcal V^{-1}} = \mathcal Ve^{\pm\sqrt{\mathcal
            D}t}\mathcal V^{-1}, 
\end{equation}
so we would like one of the diagonal terms of $\sqrt{\mathcal D}$, 
i.e. an eigenvalue of $\sqrt{\bar{\mathcal B}}$, to be larger than 
$\gamma/2$ in magnitude. Hence, $\bar{\mathcal B}$ must have an
eigenvalue greater than $\gamma^2/{4}$ for the LE to be positive.

By definition, $\bar{\mathcal B} = \gamma^2/4 - \bar{\mathcal A}$, so
if $\lambda_B$ is an eigenvalue of $\bar{\mathcal B}$, then $\lambda_A
= \gamma^2/4 - \lambda_B$ is an eigenvalue of $\bar{\mathcal
  A}$. Since we need $\lambda_B > \gamma^2/4$ for the LE to be
positive, this means we need $\lambda_A < 
0$. However, the system is in thermal equilibrium, so
\begin{align}\label{ergdist}
		\overline{\partial^2_{x_ix_i}U(\mathbf x_0(t))} &=
                C\int_\Sigma \partial^2_{x_ix_i}U(\mathbf x)e^{-U(\mathbf
                  x)/D_i}\,dV \nonumber\\ 
		&= C\int_S\partial_{x_i}U(\mathbf x)e^{-U(\mathbf
                  x)/D_i}\,d\sigma  \nonumber\\
&\quad + \frac{C}{D_i}\int_\Sigma \partial_{x_i}U(\mathbf
x)^2e^{-U(\mathbf x)/D_i}\,dV, 
\end{align} 
where $C$ is the normalization constant, $S$ the (fluctuating)
constant-temperature hypersurface, and $\Sigma$ the region enclosed by
$S$. Because we assumed $\bar{\mathcal B}$, and hence $\bar{\mathcal
  A}$, is constant, the LHS of Eq.~\eqref{ergdist} is by definition just
$A_{ii}$. Moreover, $x_i$ is bounded by $S$, so increasing $x_i$ for
any $i$ will lead to higher potential energy, which means
$\partial_{x_i}U(\vec x)$ is nonnegative on $S$. It follows the RHS of
Eq.~\eqref{ergdist} is nonnegative, so $A_{ii} \geq 0$. Consequently,
all the diagonal entries of $A$ must be positive, and in particular
$\tr \bar{\mathcal A} \geq 0$.

It is now clear why the LE is always nonpositive in the 1-dimensional
case, even with an external noise drive. In the 1-dimensional case,
$\bar{\mathcal A}$ is a scalar, so it is trivially true that $\bar{\mathcal A} =
\tr\bar{\mathcal A} = \lambda_A$. However, as stated above, in order for the
LE to be positive, we need $\lambda_A < 0$, while from
Eq.~\eqref{ergdist}, $\tr \bar{\mathcal A} \geq 0$. Hence, there are simply
not enough degrees of freedom in one dimension to satisfy these two
conditions simultaneously and produce chaos. Of course, this result is
not entirely surprising since it is well known that 1-dimensional
Hamiltonian systems are not chaotic (although driven ones can
be). Nonetheless, the calculation shows that in 1-dimensional systems
with noise-induced chaos, either the noise source is not external, or
the noise is such that the assumptions we made no longer hold, e.g.,
lack of a fluctuation-dissipation relation.

On the other hand, let us consider a 2-dimensional system. Then the
eigenvalues of $\bar{\mathcal A}$ are
\begin{equation}\label{lambdasol}
	\lambda_A = \frac{\tr\bar{\mathcal A} \pm \sqrt{(\tr\bar{\mathcal A})^2 -
            4\det\bar{\mathcal A}}}{2}. 
\end{equation}
In order to have a positive LE, one of the solutions must be
negative. This means it is necessary for either $\tr\bar{\mathcal A} < 0$
or $\det\bar{\mathcal A} < 0$. We cannot have $\tr\bar{\mathcal A} < 0$ by
Eq.~\eqref{ergdist}, but we can choose $\det\bar{\mathcal A} < 0$. A simple
numerical illustration of this is
\begin{equation}\label{example}
  \bar{\mathcal A} = \begin{pmatrix} 
  	1 & 2 \\ 
   	2 & 1
	\end{pmatrix}.
\end{equation}
With the choice $\gamma = 2$, indeed one of the exponents in
Eq.~\eqref{Delta1sol} is positive. We would like to emphasize though
that the entries in the above matrix $\bar{\mathcal A}$ are not the actual
second derivatives of the potential; rather, they are the
\emph{time-averaged} values. If we naively treat the entries of
$\bar{\mathcal A}$ as just the second derivatives of $U(\mathbf
x_0(t))$, then the potential is the quadratic potential $U(x_1,x_2) =
(1/2)x_1^2 + (1/2)x_2^2 + 2x_1x_2$, which is unbounded below along the
line $x_1=-x_2$ and thus obviously chaotic (in an unbounded
sense). However, if the entries in $\bar{\mathcal A}$ are the
time-averaged values of $\partial_{x_i,x_j}U(\mathbf x_0(t))$ for a
particular unknown bounded potential, then the LE is positive for a
system with weak noise. 

We conclude this section with an explicit expression for the LE. We
showed that the solution to the linearized version of the Langevin
equation is given by Eq.~\eqref{Delta1sol}: 
\begin{eqnarray}\label{Deltasol2}
	{\mathbf \Delta}(t) &\sim& \cosh(\sqrt{\bar{\mathcal B}}t)e^{-\gamma t/2}
        {\mathbf v}_1(0) \nonumber\\
&&+ \sqrt{\bar{\mathcal B}}^{-1}\sinh(\sqrt{\bar{\mathcal B}}t)e^{-\gamma t/2}
{\mathbf v}_2(0). 
\end{eqnarray}
where ${\mathbf v}_1(0) = {\mathbf\Delta}(0)$ and ${\mathbf v}_2(0) =
\dot{\mathbf\Delta}(0) + ({\gamma}/{2}){\mathbf\Delta(0)}$, and
$\bar{\mathcal B}$ is given by Eq.~\eqref{zlin}. Note every eigenvalue
of $\bar{\mathcal B}$ is real as it's a real symmetric  matrix, so
every eigenvalue of $\sqrt{\bar{\mathcal B}}$ is purely real or purely
imaginary. Let $\lambda_{\sqrt{B}+}$ be 
the maximum \emph{real} eigenvalue of $\sqrt{\bar{\mathcal B}}$, and
$\lambda_{\sqrt{B}-}$ be the minimum \emph{real} eigenvalue of
$\sqrt{\bar{\mathcal B}}$, and let $\lambda_{max} =
\max(|\lambda_{\sqrt{B}+}|,|\lambda_{\sqrt{B}-}|)$. Then we have the
following cases:
\begin{enumerate}
	\item If $\lambda_{max} > {\gamma}/{2}$, then
          ${\mathbf\Delta}(t) \sim e^{(\lambda_{max}-\gamma/2)t}$ for large
          $t$, so the LE is positive. 
	\item If $\lambda_{max} = {\gamma}/{2}$, then
          ${\mathbf\Delta}(t) \sim K$ for some constant $K$, so the LE is
          0. 
	\item If $\lambda_{max} < {\gamma}/{2}$ or all the eigenvalues
          are imaginary, then ${\mathbf\Delta}(t) \sim 
          e^{(\lambda_{max}-\gamma/2)t}$ for large $t$, so the LE is
          negative. 
\end{enumerate}
As we can see from above, the damping coefficient is ``suppressing''
chaos, since if we take $\gamma \to 0$, then a system with a very
small positive $\lambda_{max}$ is chaotic. But if we slightly increase
$\gamma$ and presume the long time average $\bar{\mathcal A}(t)$, and
hence $\lambda_{max}$ is not affected, then the LE is smaller in this
new system. We remark that for external noises, the noise term is
decoupled from the damping mechanism. Therefore, the damping mechanism
is associated just with the system, so the system is Hamiltonian only
if the damping coefficient $\gamma$ is zero. In this case, the third
possibility given above is impossible as $\lambda_{max} \geq 0$ by
construction. As our Hamilitonian system remains Hamilitonian after
coupling it to a Hamilitonian heat bath, this shows that our
conclusion is consistent with the fact that for Hamilitonian systems,
there are equal number of positive and negative LEs in the spectrum,
so the maximal LE is always nonnegative, i.e. only the first and
second options above are valid.

\section{Systems with Internal Noise}
\label{int_noise}

We now proceed to examine internal noise. As before, the Langevin
equation is 
\begin{equation}\label{langeq3}
	\ddot{\mathbf x}(t) + \gamma\dot{\mathbf x}(t) =
        -{\mathbf\nabla} U({\mathbf 
        x}(t)) + \sqrt{\mathcal D}\,dW_{int}(t), 
\end{equation}
where the only difference from Eq.~\eqref{langeq} is the fact that
$dW_{int}(t)$ is an additive internal noise term. (In principle, the
coupling can be more complicated, but we ignore that here.) The noise
term arises from coupling the system self-consistently to an external
dynamic structure. As discussed in the Introduction, these degrees of
freedom could be a dynamical model for a heat bath, such as a
collection of harmonic oscillators, or more generally, `fast' modes in
a system coupled to slower modes of physical interest. In any case,
let us suppose there are $m$ such outside structures. Then we can
label the position and momentum variables of these outside structures
as ${\mathbf q}(t) = (\mathbf q_1(t),\ldots,\mathbf q_m(t))^T$ and
${\mathbf p}(t) = (\mathbf p_1(t),\cdots,\mathbf p_m(t))^T$,
respectively. Note that each $\mathbf q_i$ and $\mathbf p_j$ is an
$n$-dimensional vector representing the $n$-dimensional space the
system is in. As always, we remember that ${\mathbf x}$ is an
$n$-dimensional column vector, $\gamma$ is a scalar, and $\mathcal D$
is an $n \times n$ diagonal matrix, with the $k$th diagonal entry,
denoted $D_k$, being the noise intensity along $x_k$.

We now emphasize that since we have internal noise, the term
$dW_{int}(t)$ is inherently dependent on both the system initial
conditions, $\mathbf x(0)$ and $\dot{\mathbf x}(0)$, and the `noise'
initial conditions, $\mathbf q(0)$ and $\mathbf p(0)$. This time, when
we linearize the equation, we must decide to either only perturb the
system initial conditions or those for both the system and noise. But
perturbing the system initial conditions is the same as perturbing
both the system and the noise as they are coupled, therefore, it
doesn't matter which one we choose. This does not mean that coupling a
system to a noise source cannot change the system's Lyapunov
exponent. Rather, we are stating that for a system already coupled to
an internal noise source, the exponent obtained by perturbing the
system initial conditions is the same as that obtained by perturbing
the initial conditions of both the system and the noise. For the rest
of the section we will perturb initial conditions for both the system
and noise during the linearization process.

Next, we note that here `noise' is a term we use for a complicated
underlying process with unknown exact behavior. Hence, for a
particular internal Gaussian white noise realization $dW_{int}(t)$, we
can write
\begin{equation}\label{explicitnoise}
	\sqrt{\mathcal D}\,dW_{int}(t) = \sum_i {\mathbf h}_i({\mathbf
          p}(0),{\mathbf q}(0),{\mathbf x}(0),\dot{\mathbf x}(0),t), 
\end{equation}
where $\mathbf h_i$'s are unknown $n$-dimensional column vectors
involving the initial conditions, such that when averaged over noise
realizations (denoted as $\langle \cdots \rangle_n$), 
\begin{eqnarray}\label{whitenoisereq}
	\langle \sqrt{D_i}\,dW_{int}(t) \rangle_n &=& 0, \nonumber\\
\langle \sqrt{D_i}\,dW_{int}(t)\sqrt{D_j}\,dW_{int}(t') \rangle_n &=&
\sqrt{D_iD_j}\delta(t-t') \nonumber\\
\end{eqnarray}
for all $i,j$. Now when we linearize Eq.~\eqref{langeq3}, we perturb
the initial values ${\mathbf x}(0)$, $\dot{\mathbf x}(0)$, ${\mathbf
 q}(0)$, and ${\mathbf p}(0)$. Denote the perturbed variables
${\mathbf\Delta}(t)$, $\dot{\mathbf \Delta}(t)$,
${\mathbf\Delta}_q(t)$, and ${\mathbf\Delta}_p(t)$, respectively, and
let ${\mathbf x}_0(t)$ be the fiducial trajectory. Hence, the
linearized equation is
\begin{equation}\label{intlineq}
	\ddot{\mathbf\Delta}(t) + \dot{\mathbf\Delta}(t) =
        -\mathcal A(t){\mathbf \Delta}(t) +
        \delta(\sqrt{\mathcal D}\,dW_{int}(t)), 
\end{equation}
where $\mathcal A(t) \equiv (\partial_{x_ix_j}U({\mathbf x}_0(t)))$,
the Jacobian of ${\mathbf \nabla} U({\mathbf x}_0(t))$, and 
\begin{equation}\label{noisepert}
	\delta(\sqrt{\mathcal D}\,dW_{int}(t))= \sum_i {\mathbf
          \Pi}_i(\tilde\Delta) 
\end{equation}
is the perturbation of the noise term. Here, $\tilde\Delta = (I,t)$,
where $I$ denotes the initial conditions for the system and the noise
such as $\mathbf \Delta _p(0)$ and ${\mathbf p}(0)$. Note that every
term in ${\mathbf \Pi}_i$ must, to first order (i.e. we Taylor expand
functions to first order), be proportional to one of the perturbed
initial conditions, as all the terms not proportional to an initial
condition have canceled out from the linearization. Furthermore, note
that once we linearized the equation, we no longer have a stochastic
ODE. The reason is because our perturbations are exact, so even if in
Eq.~\eqref{langeq3}, we only knew the distribution of ${\mathbf q}(0)$
and ${\mathbf p}(0)$, in Eq.~\eqref{intlineq}, we chose exact
perturbations of the initial conditions. This means that
$\delta(\sqrt{\mathcal D}dW_{int}(t))$ is not white noise!

Now, we don't actually know the exact forms of the
${\mathbf\Pi}_i$'s, and without knowledge of these functions, we
cannot solve the linearized ODE Eq.~\eqref{intlineq}. Therefore,
rather than letting the perturbations be exact, let us allow the
perturbations to be a distribution such that $\delta(\sqrt{\mathcal
 D}\,dW_{int}(t))$ is Gaussian white noise when averaged over all
possible perturbations. In other words, we choose the distribution to
satisfy for all $i,j,l,m$
\begin{align}
	\langle \Pi_{il}(\tilde\Delta) \rangle_p &= 0, \label{pertcond1}\\
	\sum_{i,j}\langle
        \Pi_{il}(\tilde\Delta)\Pi_{jm}(\tilde\Delta) \rangle_p &=
        \sqrt{K_lK_m}\delta(t-t') \label{pertcond2},
\end{align}
where $\langle\cdots\rangle_p$ denotes averaging over perturbations
(so we can pull ${\mathbf x}(0),\dot{\mathbf x}(0),{\mathbf q}(0)$ and
${\mathbf p}(0)$ out as they don't depend on noise
perturbations). Here $\Pi_{il}$ is the $l$th component of ${\mathbf
 \Pi}_i$, and $\mathcal K$ is the diagonal matrix analogous to
$\mathcal D$. While the above two conditions can certainly be
satisfied if the coupling is linear (so ${\mathbf h} \rightarrow
{\mathbf\Pi}$ by changing ${\mathbf x}(0),\dot{\mathbf x}(0),{\mathbf
 q}(0),{\mathbf p}(0)$ to their perturbed quantities) by virtue of
Eq.~\eqref{whitenoisereq}, it is unclear as to whether the above two
conditions can be satisfied for all general couplings between system
and noise.

We should point out that it is fine to take the perturbations along
some directions to be zero, since perturbing any one initial condition
causes the trajectory to be perturbed in all dimensions, as long as
the variables are coupled. We will henceforth work under the
assumption that Eqs.~\eqref{pertcond1} and \eqref{pertcond2} can be
satisfied in the system we're working in. In particular, note that
${\mathcal K} \rightarrow 0$ as ${\mathbf\Delta}(0),
\dot{\mathbf\Delta}(0), {\mathbf\Delta}_q(0), {\mathbf\Delta}_p(0)
\rightarrow 0$ by Eq.~\eqref{pertcond2} and the fact that every term
in ${\mathbf\Pi}_i$ is proportional to one of the perturbed initial
conditions by the paragraph under Eq.~\eqref{noisepert}. Thus, the
distribution of the perturbations can be thought of as a white noise
with infinitesimal intensity matrix $\mathcal K$. This means that when
studying systems with internal noise, when we perturb the fiducial
trajectory, the noise term in it is not simply a new noise realization
of intensity $\sqrt{\mathcal D}$, as done in Ref.~\cite{broeck};
rather, it has an infinitesimal intensity.

By viewing the perturbations as distributions, we can examine the mean
and variance of the divergence behavior of perturbed trajectories due
to slightly different perturbations. Note that for calculating the
LE of a system, we don't care what the perturbation is,
as long as it's infinitesimal and not specified only in one direction
(the latter requirement makes sure we have a generic perturbation so
the direction of the maximum exponent is perturbed); this is Osledec's
Theorem. We emphasize that there is nothing special about viewing the
initial perturbation as a distribution. We could have done the same in
a noise-free system or a system with external noise. However, in both
of these cases, to do so wouldn't change anything, as the linearized
equation is Eq.~\eqref{1dlin}
\begin{equation}\label{linlangeq}
	\ddot {\vec\Delta}(t) + \gamma\dot{\vec\Delta}(t) = -\mathcal
        A(t)\vec\Delta(t), 
\end{equation}
where the initial conditions do not appear at all. It is true that the
solution contains the initial conditions, so treating the initial
perturbation as a distribution will now cause the solution to be a
distribution; nonetheless, the exponential in the solution, and hence
the LE, remains the same. Viewing the initial conditions as a
distribution only helps if our system has internal noise, in which
case we do not know all the terms in the linearized equation
Eq.~\eqref{intlineq} and hence cannot evaluate it. Yet, because the
initial perturbation is generic, letting it be a distribution does not
change the LE of our system. Therefore, by examining
the exponential behavior of the mean and standard deviation of each
variable $x_i$, we can obtain the exact LE of the
system with internal noise. The reason we cannot obtain the exponent
through only the mean is because the mean divergence is effectively
obtained by choosing a particular initial perturbation, which may not
have been a generic perturbation and hence may not give the maximum
exponent; the standard deviation, however, takes into account all
generic perturbations.

From our arguments above, the linearized equation for our system with
internal noise is 
\begin{equation}\label{intnoiselin}
	\ddot{\mathbf\Delta}(t) + \gamma\dot{\mathbf \Delta}(t) =
        -\mathcal A(t){\mathbf\Delta}(t) + \sqrt{\mathcal K}\,dW(t), 
\end{equation}
where $\mathcal A(t) \equiv (\partial_{x_ix_j}U(x_0(t)))$ as before
and $dW(t)$ is our perturbation distribution, which is Gaussian white
noise. Again let
\begin{equation}\label{zdef2}
	{\mathbf z}(t) = {\mathbf\Delta}(t)e^{\gamma t/2}, 
\end{equation}
so substitution yields
\begin{align}\label{intnoiselin2}
	&\ddot{\mathbf z}(t) - \mathcal B(t){\mathbf z}(t) =
        \sqrt{\mathcal K}\,dW(t)e^{\gamma t/2}, \nonumber\\
        \text{where}\quad & \mathcal B(t) = \frac{\gamma^2}{4} - \mathcal A(t). 
\end{align}
As before, let 
\begin{equation}\label{vdef2}
	{\mathbf v}_1(t) = {\mathbf z}(t), \quad {\mathbf v}_2(t) =
        \dot{\mathbf z}(t), 
\end{equation}
and denote ${\mathbf v}(t) = ({\mathbf v}_1(t),{\mathbf
  v}_2(t))^T$. Then our second order ODE becomes 
\begin{align}\label{veq2} 
	\frac{d}{dt}\begin{pmatrix} 
  	{\mathbf v}_1(t) \\ 
   	{\mathbf v}_2(t)
	\end{pmatrix} - \mathcal M(t) \begin{pmatrix} 
  	{\mathbf v}_1(t) \\ 
   	{\mathbf v}_2(t)
	\end{pmatrix} &= \begin{pmatrix} 
  	0 \\ 
   	\sqrt{\mathcal K}\,dW(t)e^{\gamma t/2}
	\end{pmatrix} \nonumber\\
	&\equiv {\mathbf\Gamma}(t),
\end{align}
where
\begin{equation}\label{Mdef}
	\mathcal M(t) = \begin{pmatrix} 
  	\mathbf 0 & \mathcal I \\ 
   	\mathcal B(t) & \mathbf 0
	\end{pmatrix}.
\end{equation}
We keep in mind that $\mathcal M(t)$ is in fact a $2n \times 2n$
matrix, and ${\mathbf\Gamma}(t)$ is a $2n$-dimensional column
vector. As ${\mathbf\Gamma}(t)$ is Gaussian white noise, for all $i,j$, 
\begin{equation}\label{multnoisecond}
	\langle \Gamma_i(t) \rangle_p = 0, \quad \langle
        \Gamma_i(t)\Gamma_j(t') \rangle_p = U_{ij}(t,t')\delta(t-t'), 
\end{equation}
where
\begin{align}\label{Udef}
	\mathcal U(t,t') &= \begin{pmatrix} 
  	\mathbf 0 & \mathbf 0 \\
   	\mathbf 0 & \mathbf\Xi(t,t') \\
	\end{pmatrix}, \nonumber\\
        \mathbf\Xi(t,t') &= \begin{pmatrix} 
  	K_1 & \sqrt{K_1K_2} & \cdots &
        \sqrt{K_1K_n} \\ 
   	\sqrt{K_2K_1} & K_2 & \cdots &
        \sqrt{K_1K_n} \\ 
   	\vdots & \vdots & \ddots & \vdots \\
   	\sqrt{K_nK_1} & \sqrt{K_nK_2} & \cdots & K_n 
	\end{pmatrix} e^{\gamma(t+t')/2}.
\end{align}
Here, $\mathcal U$ is a $2n \times 2n$ matrix, with $U_{ij}(t,t') =
U_{ji}(t,t')$, so all the conditions for a multivariate Gaussian white
noise are satisfied~\cite{risken}. 

We now invoke the assumption that the trajectories are sampling an
equilibrum distribution over long times, so that $\mathcal B(t) \equiv
\mathcal B$ and thus $\mathcal M(t) \equiv \mathcal M$ are
constants. To solve Eq.~\eqref{veq2}, we first determine the
homogeneous solution ${\mathbf v}^h(t)$:
\begin{align}
	&\frac{d}{dt} {\mathbf v}^h(t) - \mathcal M {\mathbf v}^h(t) = \mathbf 0
        \notag \\ 
	\Rightarrow\quad & {\mathbf v}^h(t) = e^{\mathcal Mt} {\mathbf
          v}^h(0) \label{homeq}.  
\end{align}
We have already expanded this solution in the previous section. Next,
we want a particular solution for the inhomogeneous case of
Eq.~\eqref{veq2}. We will follow the steps given in Ref.~\cite{risken}
and write 
\begin{equation}\label{Gdef}
	\mathcal G(t) = e^{\mathcal Mt}.	
\end{equation}
Suppose $v_i^{inh}(t) = \sum_j G_{ij}(t)c_j(t)$ for some $c_j$'s. Then
\begin{equation}\label{dotinhv1}
	\dot v_i^{inh}(t) = \sum_j\left(\dot G_{ij}(t)c_j(t) + G_{ij}(t)\dot c_j(t)\right).
\end{equation}
Furthermore, by Eq.~\eqref{veq2}
\begin{align}\label{dotinhv2}
	\dot v_i^{inh}(t) &= \Gamma_i(t) + \sum_j M_{ij}v_j^{inh}(t) \nonumber\\
	&= \Gamma_i(t) + \sum_j M_{ij}G_{jk}(t)c_k(t) \nonumber\\
	&= \Gamma_i(t) + \sum_j \dot G_{ij}(t)c_j(t),
\end{align}
where we used Eq.~\eqref{Gdef} to obtain the last equality. Equating
Eqs.~\eqref{dotinhv1} and \eqref{dotinhv2} yields 
\begin{equation}\label{Geq}
	\sum_j G_{ij}(t)\dot c_j(t) = \Gamma_i(t).
\end{equation}
In matrix representation,
\begin{equation}
	\mathcal G(t)\dot{\mathbf c}(t) = {\mathbf\Gamma}(t).
\end{equation}
Thus,
\begin{eqnarray}\label{inhsol}
	\int_0^t \mathcal G(t){\dot{\mathbf c}}(t')\,dt' &=& \int_0^t \mathcal
        G(t)\mathcal G^{-1}(t')\mathcal G(t'){\dot{\mathbf
            c}}(t')\,dt' \nonumber\\ 
       &=& \int_0^t
        \mathcal G(t)\mathcal G^{-1}(t'){\mathbf\Gamma}(t')\,dt'. 
\end{eqnarray}
The LHS is just ${\mathbf v}^{inh}(t) = \mathcal G(t){\mathbf
 c}(t)$. Recalling Eq.~\eqref{Gdef}, the complete solution to
Eq.~\eqref{veq2} is 
\begin{align}\label{intnoisesol}
	{\mathbf v}(t) &= {\mathbf v}^h(t) + {\mathbf v}^{inh}(t) \nonumber\\
	&= e^{\mathcal Mt}{\mathbf v}(0) + \int_0^t e^{\mathcal
          M(t-t')}{\mathbf \Gamma}(t')\,dt'.  
\end{align}
Taking the mean of both sides and applying Eq.~\eqref{multnoisecond}
yields 
\begin{equation}\label{meansol}
	\langle {\mathbf v}(t) \rangle_p = e^{\mathcal Mt}\langle{\mathbf
        v}(0)\rangle_p, 
\end{equation}
so the mean of ${\mathbf v}(t)$ behaves in exactly the same way as in
the case for systems with external noise
(c.f. Eq.~\eqref{avgVevol}). Furthermore, we want to calculate the
variance of Eq.~\eqref{intnoisesol}. Multiplying the $i$th and $j$th
coordinate of ${\mathbf v}(t)$ using Eq.~\eqref{intnoisesol}, where
$i,j \leq 2n$, taking the average (see Ref.~\cite{risken}), and
applying Eq.~\eqref{multnoisecond}, we obtain the variance
\begin{align}\label{intnoisevar}
	\sigma_{ij}({\mathbf v}(t)) &= G_{ij}(t)\sigma_{ij}({\mathbf
          v}(0)) \nonumber\\ 
		&\quad +\sum_{k,s}\int_0^t\int_0^t
                G_{ik}(t-t_1')G_{js}(t-t_2') \nonumber\\ 
		&\quad \times
                \langle\Gamma_k(t_1')\Gamma_s(t_2')\rangle_p\,dt_1'\,dt_2'
                \nonumber\\  
	&= G_{ij}(t)\sigma_{ij}({\mathbf v}(0)) \nonumber\\
		&\quad +\sum_{k,s}\int_0^t\int_0^t
                G_{ik}(t-t_1')G_{js}(t-t_2') \nonumber\\ 
		&\quad \times
                U_{ks}(t_1',t_2')\delta(t_1'-t_2')\,dt_1'\,dt_2'
                \nonumber\\  
	&= G_{ij}(t)\sigma_{ij}({\mathbf v}(0)) \nonumber\\
		&\quad +\sum_{k,s}\int_0^t G_{ik}(t-t')G_{js}(t-t')U_{ks}(t',t')\,dt'. 
\end{align}
Hence, the variance of the $i$th variable is
\begin{align}\label{ithvar}
	\sigma_{ii}({\mathbf v}(t)) &= G_{ii}\sigma_{ii}({\mathbf
          v}(0)) \nonumber\\ 
&\quad+ \sum_{k,s}\int_0^t G_{ik}(t-t')G_{is}(t-t')U_{ks}(t',t')\,dt'.
\end{align}
Now, by definition Eq.~\eqref{Gdef}
\begin{align}
	\mathcal G(t-t') &= e^{\mathcal M(t-t')} = \exp\left[\begin{pmatrix} 
  	\mathbf 0 & \mathcal I \\
   	\mathcal B & \mathbf 0
   	\end{pmatrix} (t-t')\right] \notag\\
   	&= \begin{pmatrix} 
  	\cosh\sqrt{\mathcal B}(t-t') & \sqrt{\mathcal
          B}^{-1}\sinh\sqrt{\mathcal B}(t-t') \\ 
   	\sqrt{\mathcal B}\sinh\sqrt{\mathcal B}(t-t') &
        \cosh\sqrt{\mathcal B}(t-t') 
   	\end{pmatrix} \label{Gdifdef}.
\end{align}
This is in fact a $2n \times 2n$ matrix, so $G_{11}$ for example
\emph{does not} refer to $\cosh\sqrt{\mathcal B}(t-t')$, but to the
upper-left entry of $\cosh\sqrt{\mathcal B}(t-t')$. From
Eqs.~\eqref{zdef2} and \eqref{vdef2}, we have ${\mathbf\Delta}(t) =
e^{-\gamma t/2}{\mathbf v}_1(t)$. Let us denote $v_{1i}$ the $i$th
coordinate of ${\mathbf v}_1(t)$. Since this is also the $i$th coordinate
of ${\mathbf v}(t)$ as ${\mathbf v}(t) = ({\mathbf v}_1(t), {\mathbf
  v}_2(t))^T$, it follows that for $i \leq n$,
\begin{align}
	\sigma_{ii}(\mathbf\Delta(t)) &= \langle (\Delta_i(t) -
        \overline{\Delta_i(t)})^2 \rangle_p \notag\\ 
	&= \langle e^{-\gamma t} (v_{1i}(t) - \overline{v_{1i}(t)})^2
        \rangle_p \notag\\ 
	&= \langle e^{-\gamma t} (v_i(t) - \overline{v_i(t)})^2
        \rangle_p \notag\\ 
	&= e^{-\gamma t}\sigma_{ii}(v(t)) \label{varDelta}.
\end{align}
It follows for $i \leq n$, the variance of the $i$th component of
${\mathbf\Delta}(t)$ is 
\begin{align}\label{varDelta2}
	&\sigma_{ii}({\mathbf\Delta}(t)) = G_{ii}\sigma_{ii}({\mathbf
          v}(0))e^{-\gamma t} \nonumber\\  
	&\quad + e^{-\gamma t}\sum_{k,s}\int_0^t
        G_{ik}(t-t')G_{is}(t-t')U_{ks}(t',t')\,dt', \nonumber\\ 
\end{align}
where $\mathcal G(t-t')$ is determined by Eq.~\eqref{Gdifdef} and
$\mathcal U(t',t')$ is determined by Eq.~\eqref{Udef}. The standard
deviation of the $i$th component of $\mathbf\Delta$ is thus 
\begin{align}\label{sdDelta}
	&\sd_{ii}(\Delta(t)) = \big[G_{ii}\sigma_{ii}({\mathbf
          v}(0))e^{-\gamma t} \nonumber\\ 
&\quad + \sum_{k,s}e^{-\gamma t}\int_0^t
G_{ik}(t-t')G_{is}(t-t')U_{ks}(t',t')\,dt'\big]^{1/2}.  
\end{align}
Let us use our previous notation, where $\lambda_{\sqrt{B}+}$ and
$\lambda_{\sqrt{B}-}$ are the maximum and minimum real eigenvalues of
$\sqrt{\mathcal B}$, respectively, and $\lambda_{max} =
\max(|\lambda_{\sqrt{B}+}|,|\lambda_{\sqrt{B}-}|)$. Recall from
Eq.~\eqref{expsqrtBt} that 
\begin{equation}
	e^{\pm\sqrt{\mathcal B}t} = e^{\pm\mathcal V\sqrt{\mathcal
            D}t\mathcal V^{-1}} = \mathcal Ve^{\pm\sqrt{\mathcal
            D}t}\mathcal V^{-1}. 
\end{equation}
Suppose $\lambda_{max}$ is the $i$th entry of the diagonal matrix
$\sqrt{\mathcal D}$. Then the $i$th row of $e^{\pm\sqrt{\mathcal
    D}t}\mathcal V^{-1}$ are all entries involving $e^{\lambda_{max}}$,
where we choose the sign of $\sqrt{\mathcal D}$ so $\lambda_{max}$ is
nonnegative. Thus, every entry in $e^{\pm\sqrt{\mathcal B}t} = \mathcal
Ve^{\pm\sqrt{\mathcal D}t}\mathcal V^{-1}$ involves
$e^{\lambda_{max}}$. Cancellation may occur for some terms if
$\sqrt{\mathcal B}$ has multiple eigenvalues $\lambda_{max}$, but we do
not expect cancellation to occur for all terms. Then for large $t$,
\begin{equation}
	G_{ij} \sim e^{\lambda_{max}t},
\end{equation}
for some $i,j$. Hence, at least one term in the sum on the RHS of
Eq.~\eqref{sdDelta} is proportional to
\begin{eqnarray}
	&&e^{-\gamma t}\int_0^t e^{2\lambda_{max}(t-t')}e^{\gamma
          t'}\,dt' \nonumber\\
&&= e^{(2\lambda_{max}-\gamma)t}\int_0^t
        e^{(\gamma-2\lambda_{max})t'}\,dt' \nonumber\\ 
	&&=
        e^{(2\lambda_{max}-\gamma)t}\cdot
        \frac{1}{\gamma-2\lambda_{max}}\left(e^{(\gamma-2\lambda_{max})t}-1\right)    
        \nonumber\\ 
	&&\sim |e^{(2\lambda_{max}-\gamma)t} - 1|,
\end{eqnarray}
where we used Eq.~\eqref{Udef} to conclude $U_{ij} \sim e^{\gamma
 t'}$. Hence, this term has a larger exponential than the first term
on the RHS of Eq.~\eqref{sdDelta} and dominates the expression. It
follows that
\begin{equation}\label{sdApprox}
	\sd_{ii}({\mathbf\Delta}(t)) \sim \sqrt{|e^{(2\lambda_{max}-\gamma)t} - 1|}.
\end{equation}
Since the mean of ${\mathbf\Delta}(t)$ is the RHS of
Eq.~\eqref{Delta1sol} with different initial conditions, when
$\lambda_{max} \geq \gamma/2$, the mean and standard deviation of
${\mathbf\Delta}$ have the same exponential behavior, while if
$\lambda_{max} < \gamma/2$ or $\lambda_{max}$ does not exist (the
eigenvalues are imaginary), the constant term in the standard deviation
dominates. As in the external noise case, every eigenvalue of
$\sqrt{\mathcal B}$ is either purely real or purely imaginary, and we
then have the following cases: 
\begin{enumerate}
	\item If $\lambda_{max} > {\gamma}/{2}$, then
          $\sd_{ii}(\Delta(t)) \sim e^{(\lambda_{max}-\gamma/2)t}$ for
          large $t$, so the LE is positive. 
	\item If $\lambda_{max} = {\gamma}/{2}$, then
          $\sd_{ii}(\Delta(t)) \sim K$ for some constant $K$, so the
          LE is 0. 
	\item If $\lambda_{max} < {\gamma}/{2}$, then
          $\sd_{ii}(\Delta(t)) \sim K$ for large $t$ for some constant $K$,
          so the LE is 0. 
	\item If the eigenvalue(s) of $\sqrt{B}$ are all imaginary,
          then $\sd_{ii}{\Delta(t)}$ is sinusoidal, so the LE is 0. 
\end{enumerate}
This strongly resembles the cases in the previous section, when we
examined systems with external noise. However, we do note that the
internal noise now prevents exponential convergence from
occurring. Intuitively, this makes perfect sense, since while the
damping term $\gamma$ causes the phase space trajectories to converge,
the noise contributes energy into the system and hence counteracts the
damping. Clearly, since we are assuming the noise is weak, it cannot
have an impact on the chaotic behavior of systems with positive
LEs. However, if the system by itself has a negative exponent, then
the internal noise will begin to dominate the separation of nearby
trajectories after a long time, as the noise terms stay constant while
the system terms become ``weaker'' due to damping. We would like to
point out, however, that even with internal noise, we cannot have
1-dimensional chaos. This conclusion can be easily seen by the same
argument used for the external noise case.

\section{Conclusion}
\label{conclusion}

Our primary purpose here was to understand the basic issues in
defining the LE for a noisy system, in a situation where a controlled
analysis is possible. To do this we first provided a context where
`noise' is more or less clearly defined, by exploiting the oscillator
heat bath paradigm. Although this paradigm is by no means completely
general, it serves as an illustrative example for what, in principle,
needs to be worked out in more complex situations. By using this
model, we can define the LE in an uncontroversial way, by first
linearizing around the dynamical trajectory, and only later
considering what terms need to be thought of as noise, and under what
circumstances.

We distinguished in our work between external and internal noise to
avoid dynamical inconsistencies (see, e.g., the discussion in
Ref.~\cite{ngvk}). By setting up the the definition of noise following
the oscillator heat bath approach, we first considered the case of
external noise as an uncontrolled limit of the model. Even in this
case, the second-order nature of the equations of motion and the
existence of a fluctuation-dissipation theorem helps us to arrive at
reasonable conclusions about the behavior of the LEs --
no chaos for one-dimensional systems, but the possibility remaining
open in higher dimensions.

Turning next to the case of internal noise perturbations, we noted
that the noise forcing terms in this case cannot be set to zero after
linearization, because of the self-consistency requirement. A residual
piece remains, proportional to a set of (unknown) initial conditions
for the bath. In the case of a linear system-bath coupling we can
proceed by constructing a particular distribution of initial
conditions such that when averaged over it, the perturbations arising
from the initial conditions do have the properties of Gaussian white
noise, characterized by an infinitesimal noise intensity matrix
${\mathcal K}$. So, when the fiducial trajectory is perturbed, it is
not perturbed by the original noise strength $\sqrt{\mathcal{D}}$,
which can be much larger. The late-time limit of the standard
deviation of the perturbed trajectory ensemble can be used to find the
maximal LE, using once again the fact that the
trajectories are exploring a canonical distribution.

Although our analysis helped shed some light on the relationship
between chaos and weak noise, it is only a small step towards
understanding this complex yet fascinating relationship. For instance,
our treatment was restricted to the case of thermal equilibrium, and
we did not consider systems driven by external time-dependent
forces. In principle, explicit time-dependences can be introduced into
the oscillator models~\cite{sh_hek}, but the possible lack of a stable
late-time distribution will likely restrict the statements that can be
made on an analytical basis.

Furthermore, our analysis also raises some possible implications on how noise induced chaos may arise. Noise-induced chaos is chaotic behavior in a system that arises only when the system is coupled to noise. From our remarks in
the last paragraph of Section~\ref{int_noise}, there cannot be noise induced chaos in
1-dimensional ergodic (equilbrium) systems, so let us consider the
case of higher-dimensions. From Eq.~\eqref{zlin} and taking $\mathcal
B(t) \equiv \mathcal B$ constant, we see that $B_{ii} = \gamma^2/4 -
A_{ii}$. But we also know from Eq.~\eqref{ergdist} that 
\begin{eqnarray}\label{ergdist2}
	A_{ii} &=& C\int_S\partial_{x_i}U(x)e^{-U(\vec
          x)/D_i}\,d\sigma \nonumber\\
        &&+ \frac{C}{D_i}\int_\Sigma \partial_{x_i}U(\vec x)^2e^{-U(\vec
          x)/D_i}\,dV, 
\end{eqnarray}
so $B_{ii}$, and hence $\mathcal B$, is dependent on noise
intensity. Because the eigenvalues of $\sqrt \mathcal B$ determine the
LE of the system, this means if $\mathcal B$ changes, the LE will
potentially also change. In particular, let us suppose our system has
a $\lambda_{max}$ slightly less than $\gamma/2$. Since our analysis of
systems with external noise formally resembles that of noise-free
systems as the noise term disappears via linearization (see
Eq.~\eqref{1dlin} and the paragraph below), this means if
$\lambda_{max}$ is slightly less than $\gamma/2$, then our system has
a negative LE. Now, let us couple this system to either external or
internal noise. If the noise intensity $D_i$ is small enough such that
the system coupled to the noise is still approximately ergodic, but
large enough to shift the eigenspectrum of $\sqrt{\mathcal B}$ such
that $\lambda_{max}$ is now slightly greater than $\gamma/2$, then our
system now has a positive LE, resulting in noise-induced chaos. However, how to couple the noise so that such a shift in the eigenspectrum occurs is still an open question, and one that is worth exploring.

\section{Acknowledgements}

The work of T.H. was supported in part by a SULI award at Los Alamos
National Laboratory, and he acknowledges the many discussions he had
with Hideo Mabuchi. S.H. acknowledges past discussions with Tanmoy
Bhattacharya, Kurt Jacobs, Henry Kandrup, Elaine Mahon, Govindan
Rangarajan, Robert Ryne, Kosuke Shizume, and Robert Zwanzig.

\end{document}